\RequirePackage{fixltx2e}
\documentclass[aps,prb,reprint,groupedaddress,amsmath,amssymb]{revtex4-1}

\usepackage{graphicx}
\usepackage{subcaption}
\usepackage{dcolumn}
\usepackage{bm}
\usepackage{multirow}
\usepackage[usenames,dvipsnames,svgnames,table]{xcolor}
\usepackage{float}
\usepackage{overpic}
\usepackage{gensymb}
\usepackage{times}
\usepackage{textcomp}
\usepackage{rotating}
\usepackage{textcomp}
\usepackage{threeparttable}
\usepackage{titlesec}
\newcommand{\sub}[1]{\textsubscript{#1}}
\newcommand{\sups}[1]{\textsuperscript{#1}}

\begin{document}


\title{Giant spin-orbit splitting of point defect states in monolayer WS\sub{2}} 


\author{Wun-Fan Li}
\email[Corresponding author: ]{W.F.Li@uu.nl}
\affiliation{Soft Condensed Matter, Debye Institute for Nanomaterials Science,
Utrecht University, Princetonplein 5, 3584CC Utrecht, The Netherlands}
\author{Changming Fang} 
\affiliation{Soft Condensed Matter, Debye Institute for Nanomaterials Science,
Utrecht University, Princetonplein 5, 3584CC Utrecht, The Netherlands}
\author{Marijn A. van Huis}
\affiliation{Soft Condensed Matter, Debye Institute for Nanomaterials Science,
Utrecht University, Princetonplein 5, 3584CC Utrecht, The Netherlands}

\date{\today}

\begin{abstract}
The spin-orbit coupling (SOC) effect has been known to be profound in monolayer pristine transition metal dichalcogenides (TMDs). 
Here we show that point defects, which are omnipresent in the TMD membranes, exhibit even stronger SOC effects and change the physics of the 
host materials drastically. 
In this Article we chose the representative monolayer WS\sub{2} slabs from the TMD family together with seven typical types of point defects including 
monovacancies, interstitials, and antisites. We calculated the formation energies of these defects, and studied the effect of spin-orbit coupling (SOC)
 on the corresponding defect states. We found that the S monovacancy (V\sub{S}) and S interstitial (adatom) have the lowest formation energies. 
In the case of V\sub{S} and both of the W\sub{S} and W\sub{S2} antisites, 
the defect states exhibit giant splitting up to 296 meV when SOC is considered. 
Depending on the relative position of the defect state with respect to the conduction band minimum (CBM), the hybrid functional HSE will either increase the splitting
by up to 60 meV (far from CBM), or decrease the splitting by up to 57 meV (close to CBM).
Furthermore, 
we found that both the W\sub{S} and W\sub{S2} antisites possess a magnetic moment of 2 $\mu_{B}$ localized at the antisite W atom and the neighboring W atoms. 
All these findings provide new insights in the defect behavior under SOC point to new possibilities for spintronics applications for TMDs. 
\end{abstract}


\maketitle

\section{Introduction}
The transition metal dichalcogenides (TMDs) are a member of 
the layered 2D van der Waals (vdW) materials, in which the atoms are bound by intra-layer chemical bonding and inter-layer vdW bonding. 
Among many other TMDs, the molybdenum dichalcogenides and tungsten dichalcogenides (MX\sub{2}, M=Mo or W, and X= S, Se, or Te) are the group 6 branch of the whole 
TMD family and have attracted much scientific attention. 
Theoretically, the most stable structure of MX\sub{2} consists of one layer of transition metal atoms sandwiched by two layers of chalcogen atoms 
with a prismatic coordination, forming the so-called 1H form\cite{ataca-2012-2}.  
Due to the weak inter-layer vdW interaction, TMDs can be exfoliated from bulk into the few-layer or monolayer (ML) forms. When reducing the number of 
layers from bulk to ML, the band gap of TMDs evolves from an indirect band gap to a direct band gap with an increased gap size 
due to quantum confinement\cite{mak, kuc}. The layer-dependent tunability of the electronic structure together with other distinct physical properties of ML TMDs
make them promising candidates of applications in fields like electronics, optoelectronics, spintronics and valleytronics, 
sensing, and catalysis\cite{wang, roldan, heine, ganatra}.   

There are two effects governing the band structure (BS) of MX\sub{2}, 
namely crystal field (CF) splitting and spin-orbit (SO) splitting ($\Delta_{\text{SO}}$). These two effects strongly affect the electronic properties of MX\sub{2}
and influence in particular the $d$ bands of the transition metal. 

\begin{figure}
\includegraphics[width=\linewidth, trim=0cm 5cm 0cm 5cm, clip=true]{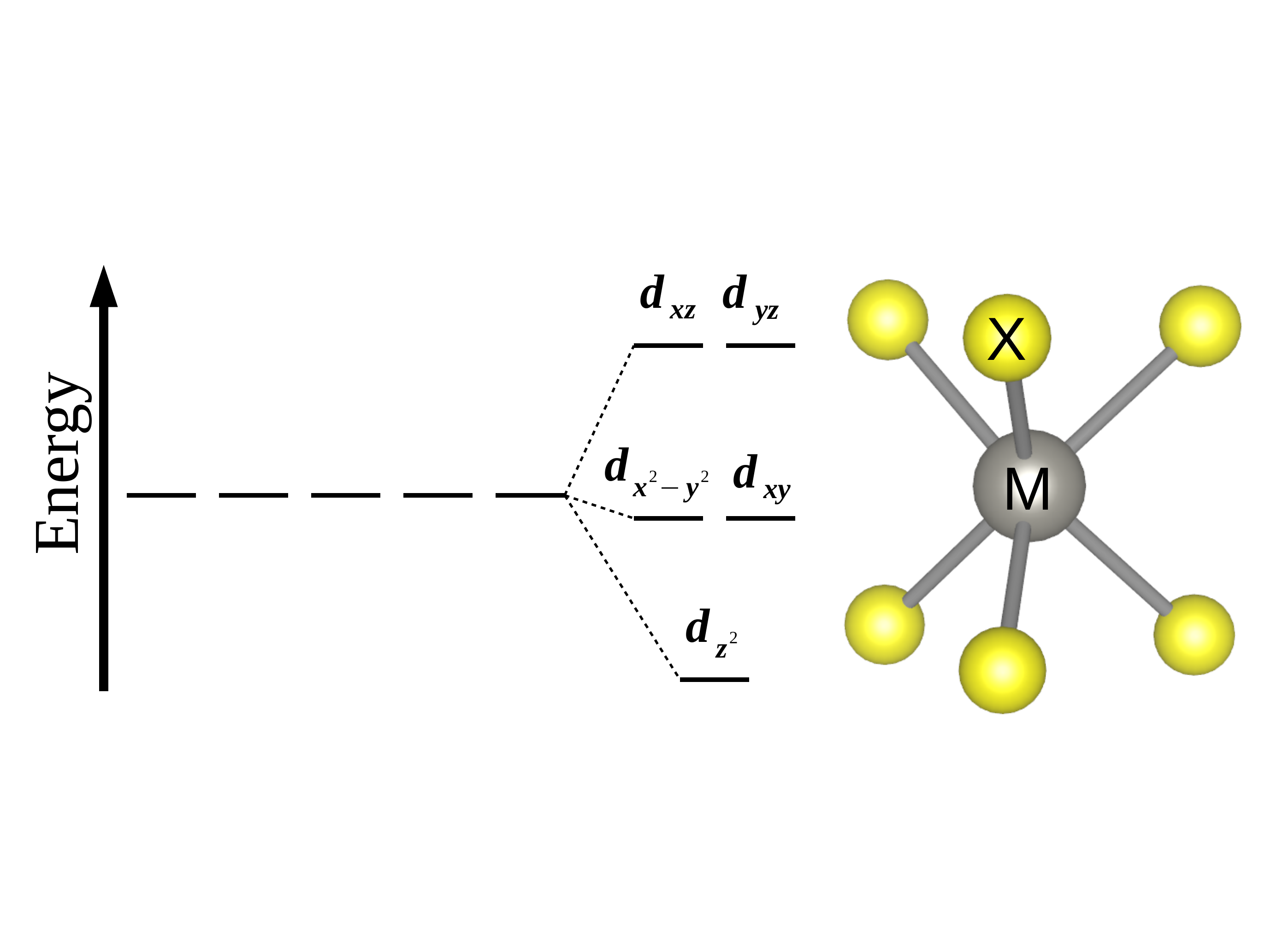}
\caption{Schematic of the energy splitting of the transition metal $d$ bands under the crystal field. The coordination is trigonal prismatic.}
\label{fig:coord}
\end{figure}

According to crystal field theory, the five formerly degenerate $d$ bands of the transition metal will split in energy if the transition metal is bonded to 
other ligands (the chalcogen atoms in our case), and the pattern of the energy splitting is dependent on the metal-ligand coordination geometries. 
For ML MX\sub{2} in the 1H phase, the transition metal is surrounded by six chalcogen atoms in a trigonal prismatic coordination (Fig. \ref{fig:coord}).
Consequently, the $d$ bands split according to their orientations - the more they are along the direction of the M-X bond, the higher in energy they will be 
due to the electron-electron repulsion with the X orbitals. As shown in Fig. \ref{fig:coord}, the $d_{z^2}$ orbital is the lowest in energy, and 
the $d_{x^2-y^2}$ and $d_{xy}$ orbitals are higher in energy. The $d_{xz}$ and $d_{yz}$ orbitals are the highest in energy\cite{chhowalla, schmidt}.
The Supplemental Material (SM) shows the decomposed band structures of both bulk and ML WS\sub{2} which illustrate the CF splitting of the $d$ bands
 (Figures S3 and S4). 
The order of increasing energy is $d_{z^2}<d_{x^2-y^2}=d_{xy}<d_{xz}=d_{yz}$ for both bulk and ML WS\sub{2}, as expected.   

The spin-orbit coupling (SOC) effect in MX\sub{2} materials has been discovered in the last few years\cite{zhu, xiao, li, kosmider}. 
In bulk MX\sub{2}, the system possesses both the space inversion symmetry ($E_{\downarrow}(\vec{k})=E_{\downarrow}(\vec{-k})$)and 
time inversion symmetry ($E_{\downarrow}(\vec{k})=E_{\uparrow}(\vec{-k})$). The net result is spin degeneracy in reciprocal space 
when no external magnetic field is present: $E_{\downarrow}(\vec{k})=E_{\uparrow}(\vec{k})$. However, in the case of ML MX\sub{2}, because of the 
lack of space inversion symmetry, the spin states are expected to split under SOC. Especially, the band splitting can be as large as 
463 meV for the valence band maximum (VBM) of ML WSe\sub{2} at the $K$ point in the first Brillouin zone\cite{kosmider}. 
Based on symmetry arguments\cite{zhu, kosmider}, for ML MX\sub{2} only the orbitals with magnetic quantum number 
$m_{l}\neq{0}$ will participate SO splitting. Furthermore, because the X atoms are rather light, their $p$ orbitals are not affected by the SOC effect. 
Lastly, as indicated in the BSs of ML WS\sub{2} in the SM (Fig. S4), 
the VBM and conduction band minimum (CBM) are dominated by the $d_{z^2}$ ($m_{l}=0$), $d_{xy}$ ($m_{l}=-2$) 
and $d_{x^2-y^2}$ ($m_{l}=2$) orbitals. As a result, only the $d_{xy}$ and $d_{x^2-y^2}$ orbitals will have the SO splitting.  

Besides the novel physical properties of pristine TMDs, atomic point defects are omnipresent in the materials.
Furthermore, adatom adsorption and doping on ML MX\sub{2} is especially achievable by virtue of their 2D surface nature. 
Both the naturally occurring and chemically or physically introduced point defects in MX\sub{2} will extensively modulate the physical properties such as   
charge transport, magnetism, optical absorption, and absorbability
\cite{ataca-2011, ma, ataca-2011-2, koh, wei, qiu, liu-apl, carvalho, chang, yuan, rastogi, lu, chow, liu-rscadv, hu, zhao, li-2}, 
thus control the applicability of the material.
The crucial role of point defects has triggered many studies to investigate their behavior in ML MX\sub{2}.  
Liu \textit{et al.} identified the atomic defects and visualized their migrations on ML MoS\sub{2}\cite{liu}. 
Komsa \textit{et al.} found that electron beam irradiation generates sulfur monovacancies (V\sub{S}) and also cause these defects to migrate and 
aggregate\cite{komsa-2012, komsa-2013}. Zhou and \textit{et al.} carried out a joint experiment and theory study and investigated several types of defects and 
their influence on the electronic structure of ML MoS\sub{2} synthesized by chemical-vapor-deposition (CVD)\cite{zhou}. 
Among the single vacancy, vacancy complexes and antisite complexes, they found that the V\sub{S} is the 
predominant point defect. First principles calculations confirmed that V\sub{S} has the lowest formation energy among all the defect kinds.
Hong \textit{et al.} did a systematic study which shows the route-dependency of predominant point defect types\cite{hong}.
In ML MoS\sub{2} synthesized by CVD and mechanical exfoliation (ME), V\sub{S} is the only dominating point defect, 
whereas in ML MoS\sub{2} fabricated by physical-vapor-deposition (PVD), the antisites Mo\sub{S2} and Mo\sub{S} are the dominant point defects. 
They also found that the Mo\sub{S} antisite possesses a local magnetic moment around the Mo defect site.
From the theoretical perspective, several exhausive works have been done to study the point defects systematically 
by virtue of the density functional theory (DFT)\cite{noh, komsa-2015-prb, halder}. Their results predict that in ML MX\sub{2}, the V\sub{S} and sulfur 
interstitial S\sub{i} have the lowest formation energy. 
 
\begin{figure*}
\includegraphics[width=\textwidth, trim=1cm 12cm 1cm 1cm, clip=true]{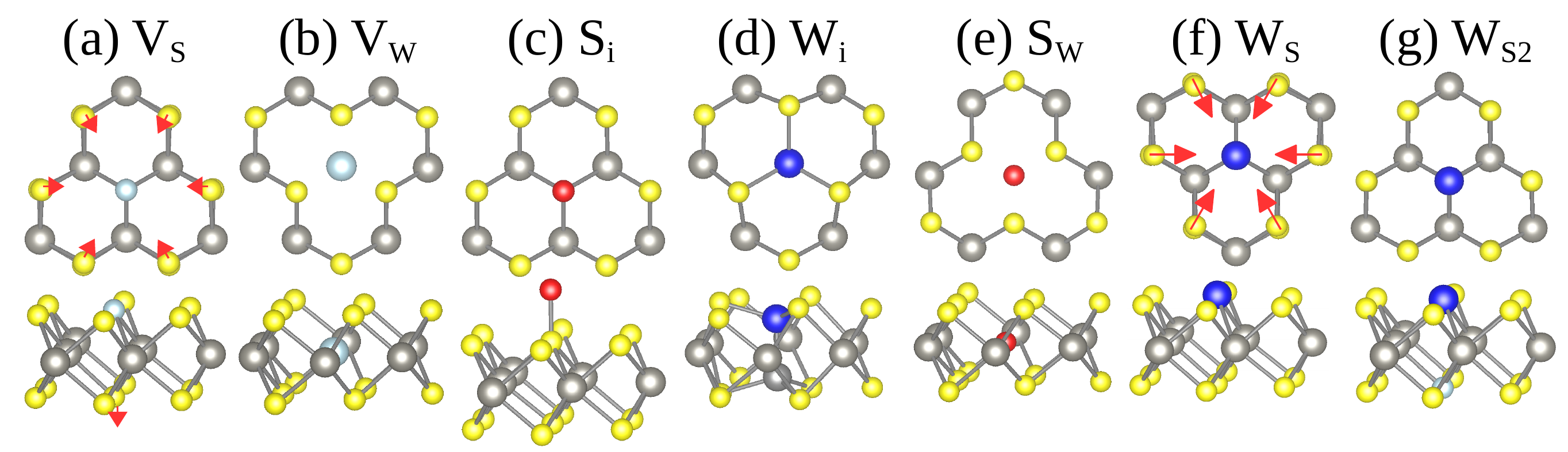}
\caption{The relaxed structures of all the defective ML WS\sub{2} supercells. 
		The vacancies are denoted by light blue circles. The defect sufur atoms are marked in red, and defect tungsten atoms in blue.  
			The arrows indicate the directions and magnitudes of the relaxations.} 
\label{fig:models}
\end{figure*}

Despite the significance of SOC and point defects for ML MX\sub{2} systems, to the best of our knowledge thus far no study has been conducted 
on the SOC effect on the electronic structure of defective ML MX\sub{2}.  
Therefore, here we investigate how the SOC effect will change the band structure (BS) of ML MX\sub{2} when different types of point defects are present. 
We chose systematically three categories of point defects: monovacancies (V\sub{S} and V\sub{W}), interstitials (S\sub{i} and W\sub{i}), and 
antisites (S\sub{W}, W\sub{S}, and W\sub{S2}). For conciseness, the ML WS\sub{2} slabs containing these defects are abbreviated as: V\sub{S}-WS\sub{2}, 
V\sub{W}-WS\sub{2}, S\sub{i}-WS\sub{2}, W\sub{i}-WS\sub{2}, S\sub{W}-WS\sub{2}, W\sub{S}-WS\sub{2}, and W\sub{S2}-WS\sub{2}, respectively.  
The relaxed structure of each point defect is shown in Figure \ref{fig:models}. 
We chose WS\sub{2} as a representative of the MX\sub{2} family as the physical and chemical properties of all the MX\sub{2} members are very similar, and 
thus the results of WS\sub{2} are expected to be applicable to other MX\sub{2} systems. 

After describing the computational settings, we will first discuss the formation energies of the selected defect species. We then chose V\sub{S}, S\sub{i}, 
W\sub{S}, and W\sub{S2} for further investigation of the SO defect state splitting. We found that 
SOC causes giant defect state splitting in the cases of V\sub{S} and W\sub{S2}, with the magnitude of the band splitting up to 194 meV for V\sub{S} and 167 meV for 
W\sub{S2} respectively. In addition, we also found that both W\sub{S} and W\sub{S2} antisites possess a magnetic moment around the antisite W atom, which 
is contrary to the previous study of MoS\sub{2}\cite{hong}. The findings in this work provide a deeper insight in the point defect physics of MX\sub{2} 
and will help developing potential  applications of MX\sub{2} in electronics and spintronics. 
   
\section{Computational details}
All calculations were performed using the DFT code VASP\cite{vasp_1, vasp_2, vasp_3} within the Projector-Augmented Wave (PAW) framework\cite{paw}. 
The exchange and correlation energies were described using the Generalized Gradient Approximation (GGA) formulated by 
Perdew, Burke and Ernzerhof (PBE)\cite{gga-pbe_1, gga-pbe_2}. 
The cut-off energy of the wave functions and the augmentation functions were 400 eV and 550 eV, respectively. 
The van der Waals correction with the optB88-vdW density functional\cite{vdw_2} was used as at the 
beginning of this study the bulk WS\sub{2} was also included\footnote{Please refer to the Supplementary Information for more details}. 
The supercell size of the ML WS\sub{2} was $6\times6$ in the $x-y$ plane, and the vacuum along the $z$ direction was larger than 16 \AA. 
These dimensions of the supercell were sufficiently large to avoid the artificial defect-defect interaction. 
A $\Gamma$-centered $2\times2\times1$ $k$-mesh was used. 
The thresholds of energy convergence and force convergence were $10^{-4}$ eV and $10^{-2}$ eV/\AA, respectively. 
We examined the SOC effect and found that it does not affect the structure but only influences the electronic properties of WS\sub{2}, 
therefore we only included SOC after geometry relaxation to obtain the band structure (BS) and DOS for the systems.  
We first performed the geometry relaxation and total energy calculation with only vdW correction included (without SOC). Then we turn on SOC, and exclude vdW 
correction for calculating the electronic properties (BS and DOS) of the relaxed geometry. Spin polarized (SP) calculations were performed for every point defect species, and only the W\sub{S} and W\sub{S2} antisites were found to be magnetic 
due to their unpaired electrons.
The initial geometry of each point defect configuration was chosen based on previous theoretical studies\cite{komsa-2015-prb, halder}.
The stringent setting described above guarantees a good convergence of defect formation energy within 0.01 eV. 

In addition to standard DFT calculations, we also performed the more advanced 
hybird functional (HSE06)\cite{hse} calculations for the defective ML WS\sub{2} which shows defect state splitting under SOC 
(the V\sub{S}-WS\sub{2}, W\sub{S}-WS\sub{2}, and W\sub{S2}-WS\sub{2}). The goal of these HSE+SOC calculations 
is to investigate how HSE will affect the defect state splitting. 
The HSE calculations were performed on the DFT-relaxed geometries and we found that HSE relaxation gave almost 
identical geometries compared to traditional DFT.  
We set the fraction of Hartree-Fock exchange functional to 0.168 by fitting the calculated band gap of ML WS\sub{2} to the experimental value. This fraction 
gives us a band gap of 2.04 eV, which is very close to the experimental value of 2.05 eV\cite{frey}.  
In the HSE+SOC calculations only the $\Gamma$ point was included as we did a test for V\sub{S}-WS\sub{2} and W\sub{S}-WS\sub{2} and found that  
a $2\times2\times2$ $k$-mesh only improves the band gap for 7 meV for V\sub{S}-WS\sub{2}, and for 13 meV for W\sub{S}-WS\sub{2}. Therefore we believe that 
$\Gamma$ is sufficient in our case. Our SO splitting of the top valence bands of perfect ML WS\sub{2} calculated by DFT is 430 meV, which is perfectly matching the 
previous DFT result of 433 meV\cite{kosmider}. HSE increases this splitting considerably to 517 meV. 
  
\section{Results and discussion}
\subsubsection{Defect formation energy}
The formation energy $E_f$ of a neutral defect is defined as\cite{chris}
\begin{equation}
    \label{eq:dfe}
    E_f=E_{defect}-E_{perfect}+\sum_{i}n_{i}\mu_{i}.
\end{equation}
In Eq. \ref{eq:dfe}, $E_{defect}$ is the total energy of the defective system, $E_{perfect}$ is the total energy of the perfect system,
$n_i$ is the number of atoms being added (plus) or removed (minus) from the perfect system, and $\mu_{i}$ is the chemical potential of the added or removed
atom. The added/removed atom is imagined to be taken from/put to an atomic reservoir, thus the energy required for creating a defect inside a system does not depend
only on the system itself, but also on the atomic resevoir, or the surrounding environment.
Chemical potentials $\mu_{i}$ are therefore needed to reflect the chemical environment surrounding the system. $\mu_{i}$ is not fixed, but a variable for which 
we could determine its boundaries by considering the formation reaction of a material. 
We give the derivation of the boundaries of $\mu_{i}$ in the case of WS\sub{2} in the SM\footnote
{Please refer to Section II in the SM for more information.}
The final expressions of the boundaries of $\mu_{W}$ and $\mu_{S}$ which are used to calculate the defect formation energies are:
\begin{subequations}
    \begin{gather}
        E_{WS_{2}}-2E_{S}\leq\mu_{W}\leq E_{W}\\
        \frac{1}{2}(E_{WS_{2}}-E_{W})\leq\mu_{S}\leq E_{S}.
    \end{gather}
\end{subequations}

The calculated defect formation energies are listed in Table \ref{tab:dfe} dependent on W-rich or S-rich chemical potentials. 

\begin{table}
\caption{Formation energies (in eV) of the defects selected in this study}
\label{tab:dfe}
\begin{ruledtabular}
\begin{tabular}{lcc}
                &W-rich &S-rich \\
    \hline
    V\sub{S}    &1.689  &2.897  \\
    V\sub{W}    &6.345  &3.928  \\
    S\sub{i}    &2.419  &1.211  \\
    W\sub{i}    &5.317  &7.733  \\
    S\sub{W}    &8.219  &4.594  \\
    W\sub{S}    &5.380  &9.005  \\
    W\sub{S2}   &6.838  &11.671 \\
\end{tabular}
\end{ruledtabular}
\end{table}
The next step is to choose relevant defect types for further study of the effect of SOC on electronic properties of the defective ML WS\sub{2} slabs.
Table \ref{tab:dfe} provides a simple criterion in terms of defect formation energy:
V\sub{S} and S\sub{i} possess the lowest formation energies in both the W-rich and S-rich conditions, thus it is sensible to select them for 
more detailed study.
Although the W\sub{S} and W\sub{S2} antisites have a higher formation energy, it has been reported that
the Mo\sub{S} and Mo\sub{S2} antisites are the predominant point defects in MoS\sub{2} synthesized by physical-vapor-deposition (PVD).
Therefore the W\sub{S} and W\sub{S2} antisites are also included in the present study\cite{hong}.
 
\subsubsection{Defect state splitings under SOC}
\begin{figure*}
	\centering
	\includegraphics[width=\textwidth]{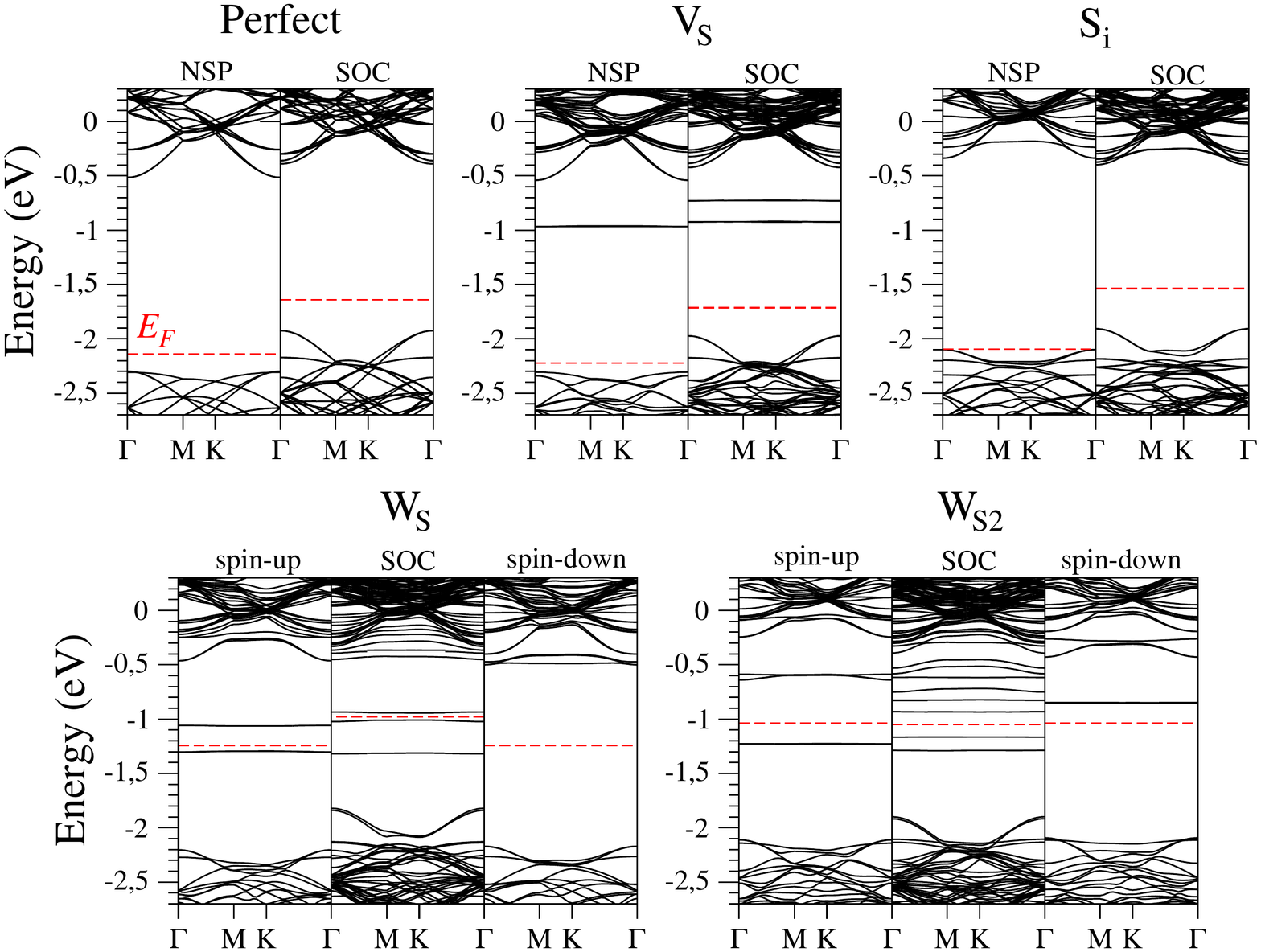}
	\caption{The band structures calculated with or without SOC for the selected WS\sub{2} slabs. NSP stands for
	non-spin-polarized non-SOC calculations, and spin-up and spin-down stand for the spin-polarized calculations, respectively. 
	Here the Fermi level is marked in red. 
	The defect state splitting can be clearly seen in the case of V\sub{S} and WS\sub{2}. However, the splitting is supressed for S\sub{i}. }
	\label{fig:band}
\end{figure*}

\begin{figure*}
	\centering
	\begin{subfigure}{0.29\textwidth}
	\includegraphics[scale=0.34]{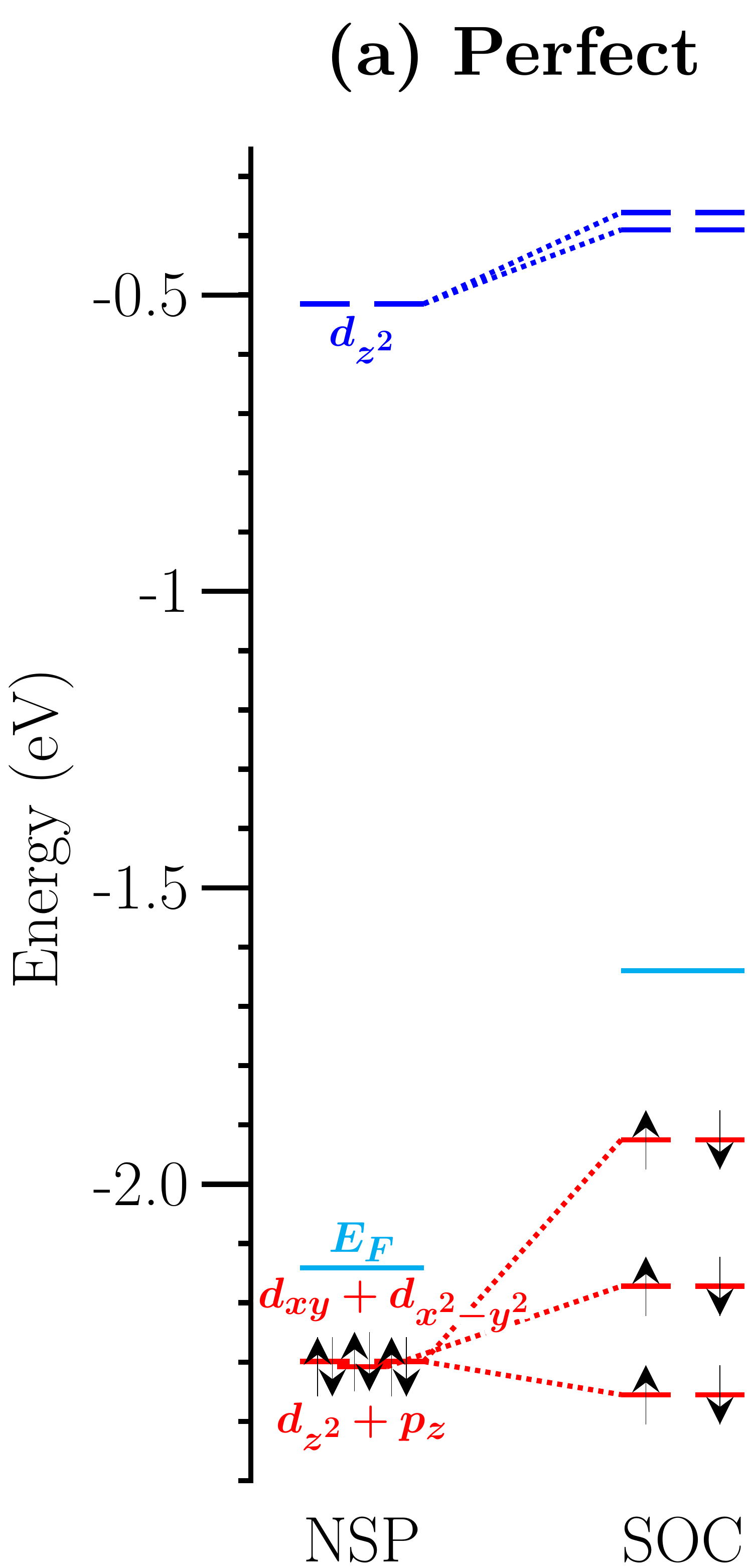}
	\end{subfigure}
	\centering
	\begin{subfigure}{0.29\textwidth}
	\includegraphics[scale=0.34]{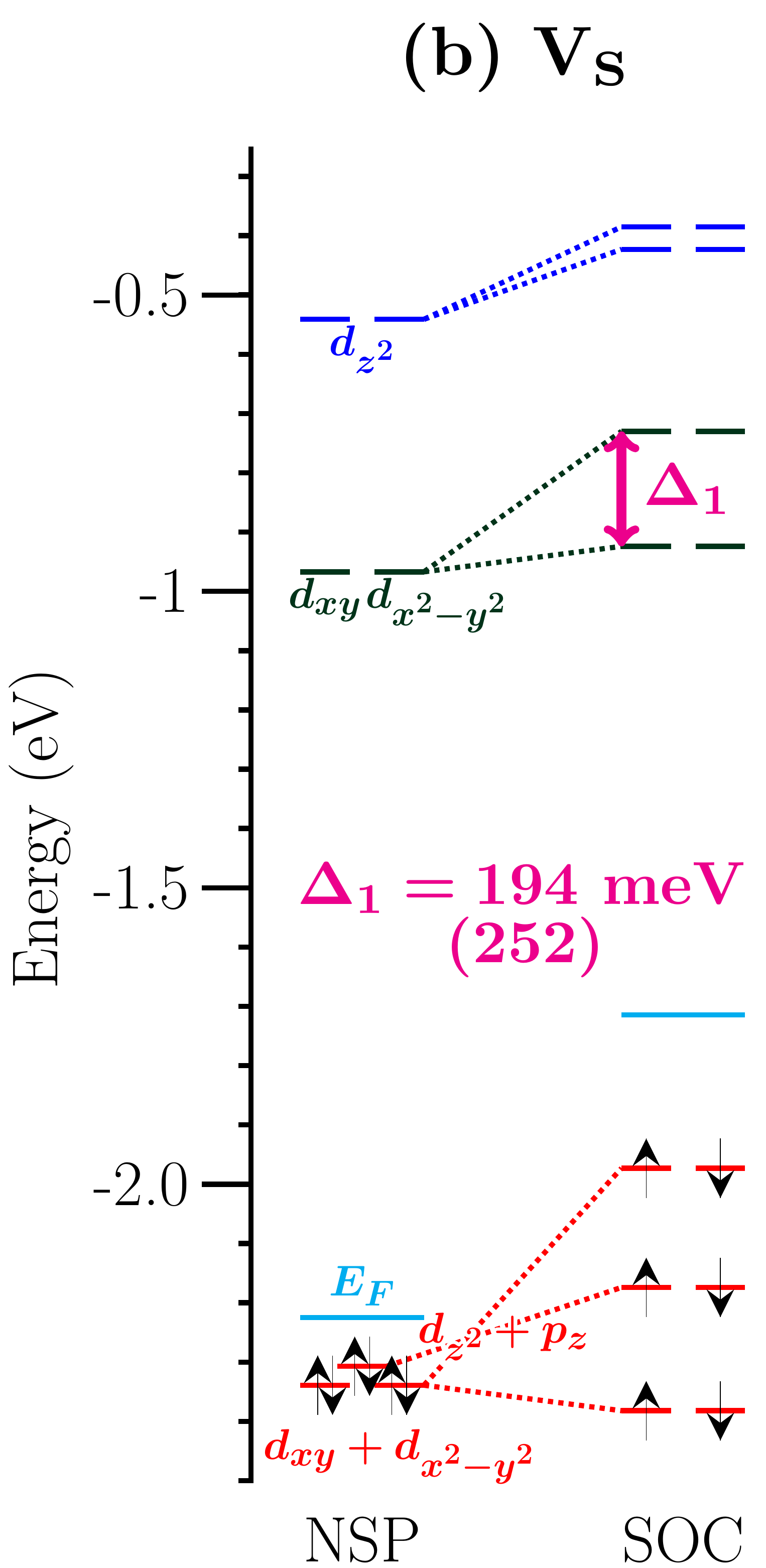}
	\end{subfigure}
	\centering
	\begin{subfigure}{0.29\textwidth}
	\includegraphics[scale=.34]{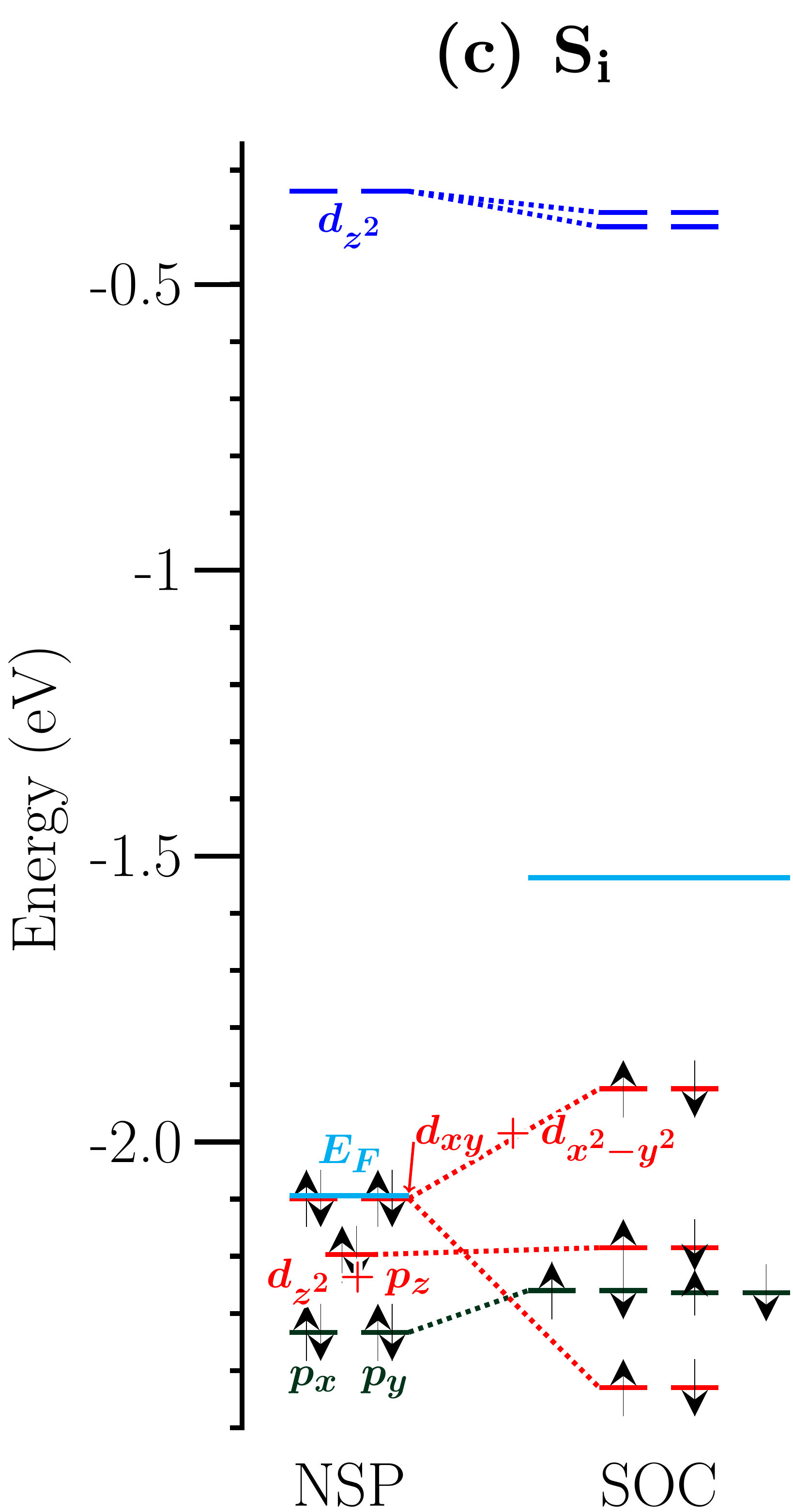}
	\end{subfigure}\\
	\vspace{.5cm}
	\centering
	\begin{subfigure}{0.42\textwidth}
	\vspace{-0.3cm}
	\includegraphics[scale=.34]{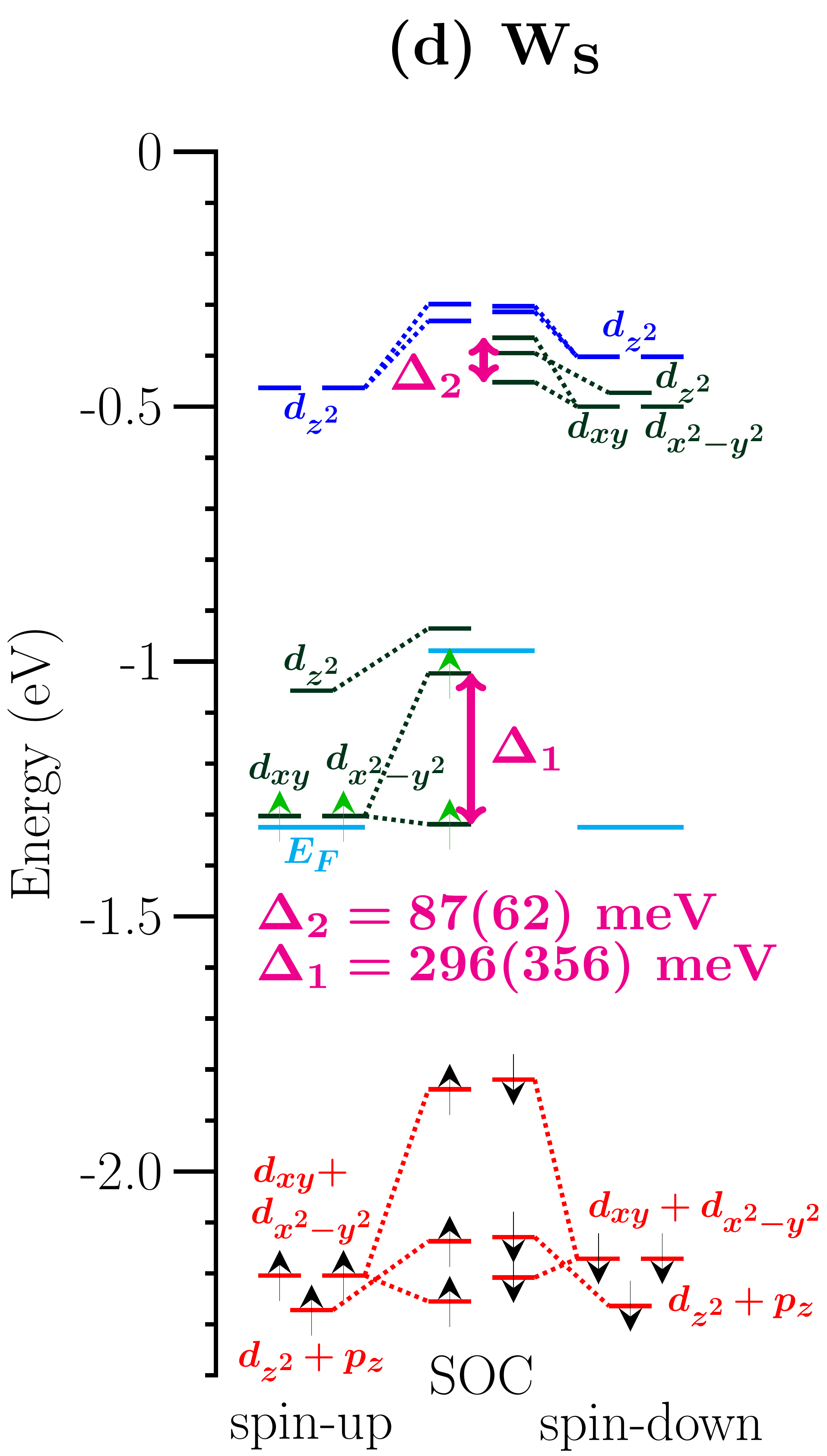}
	\label{fig:el_ws}
	\end{subfigure}
	\centering
	\begin{subfigure}{0.42\textwidth}
	\includegraphics[scale=.34]{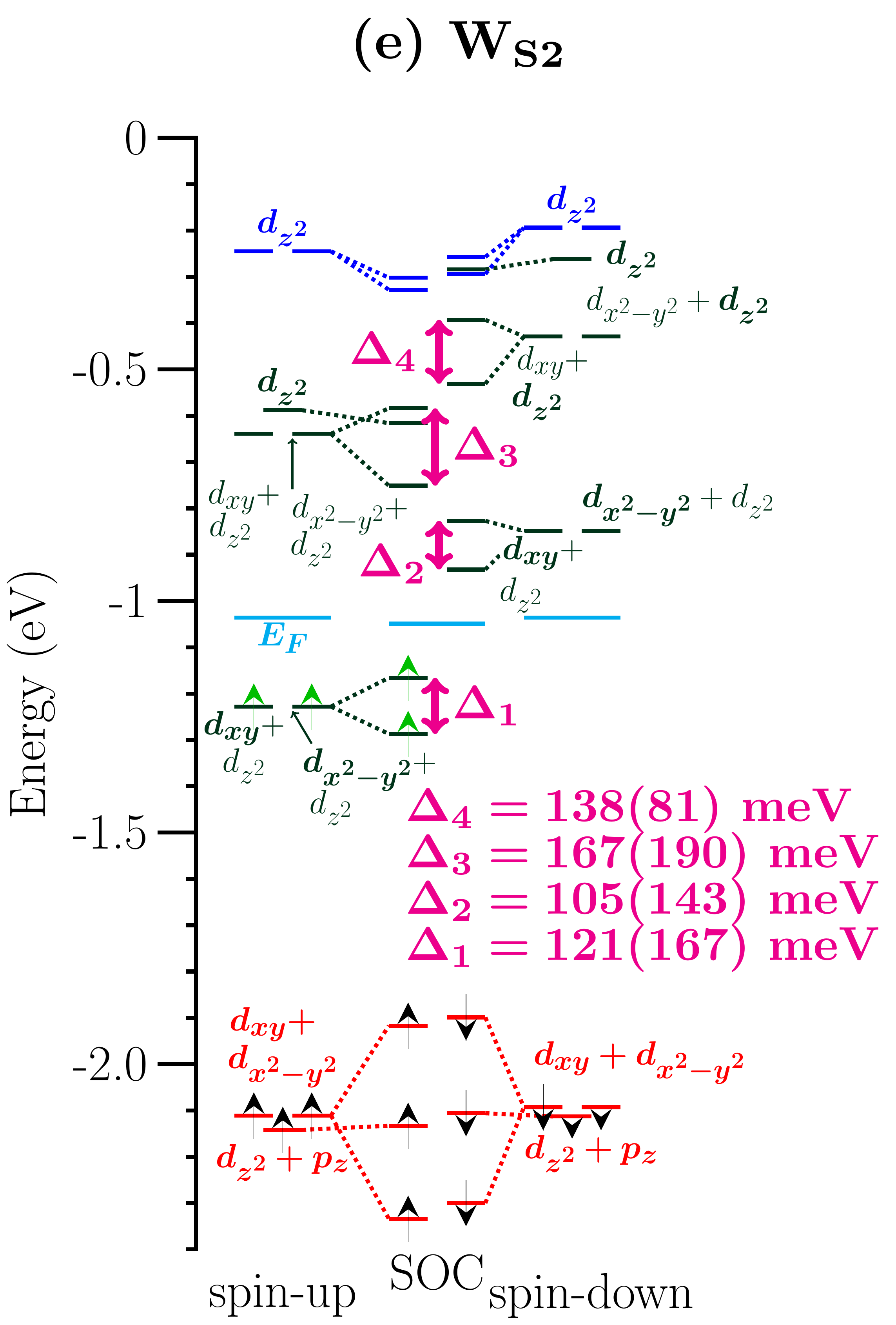}
	\label{fig:el_ws2}
	\end{subfigure}
	\caption{The energy level diagram of the WS\sub{2} systems at the $\Gamma$ point. 
	The valence bands are colored in red, defect states in green, and conduction bands in blue. 
	The Fermi level, $E_F$, is marked in cyan. 
	The electrons which contribute to magnetism for W\sub{S} and W\sub{S2} antisites are labeled in light green.
	The major orbital components of each band are indicated, where the orbitals in bold are the most predominant ones. 
	The dotted lines show the SO splittings of the energy bands. 
	The amount of the SO splitting ($\Delta$) is also shown in magenta, the values for $\Delta$in parentheses were calculated by HSE+SOC.}
	\label{fig:el}
\end{figure*}

As seen in Ref. \onlinecite{padilha} and Fig. S4 in the SM, the valence bands of MX\sub{2} are composed of the $p_{x}$ and $p_{y}$ orbitals of the X atoms (here: 
S atoms), and the $d_{xy}$, $d_{x^2-y^2}$ and $d_{z^2}$ orbitals of the M atoms (here: W atoms). 
The $d_{xz}$, $d_{yz}$ orbitals are far from the band gap region. Furthermore, Fig. \ref{fig:el} 
indicates that the top valence bands and the bottom conduction bands consist mainly of the $d$ orbitals of W atoms. 
The only $p$ orbital present is the $p_{z}$ orbital from the S atoms, and it does not split under SOC.
The calculated BSs with and without SOC are shown in Fig. \ref{fig:band}.  
We can see from Fig. \ref{fig:band} and Fig. \ref{fig:el} that irrespective of the type of point defects, 
the VBM of WS\sub{2} always splits into two bands under SOC. 
\begin{center}
	V\sub{S}
\end{center}
As discussed in the Introduction, only the W $d_{xy}$ and $d_{x^2-y^2}$ orbitals will undergo SO splitting.
This is the case for V\sub{S}. The defect states are composed of the the linear combinations of W $d_{xy}$ and $d_{x^2-y^2}$ orbitals, 
which are formerly degenerate are now split into two bands. The magnitude of the SO splitting for V\sub{S} is 194 meV.
The HSE+SOC calculation gave a SO splitting of 252 meV, which is 58 meV larger than the DFT+SOC value. 
This substantial energy difference shows the necesity of hybrid functionals in calculating the SO splitting of the defect states.    
\begin{center}
	S\sub{i}
\end{center}
In the case of S\sub{i}, the only defect state is composed of the $p_{x}$ and $p_{y}$ orbitals of the interstitial S atom, which do not split under SOC. 
This defect state is hidden in the top valence bands.  
\begin{center}
	W\sub{S}
\end{center}
For W\sub{S}, the defect states are also composed of W $d_{xy}$ and $d_{x^2-y^2}$, but they do not split when SOC is included in the calculations.
Further eigenstate analysis shows that the reason for the defect states to be kept degenerate is that the spin projections of these states in the SOC BS are 
all on the $m_{x}-m_{y}$ plane ($m_{x}$, $m_{y}$ and $m_{z}$ are the magnetization axes), in contrast to the defect states of the other three defect kinds where 
the spin projections are either mostly on along the $m_{z}$ axis (in the case of W\sub{S2}, $+m_{z}$ for spin-up and $-m_{z}$ for spin-down). 
As a result, the spin states are not split even when SOC is present. 
We performed a second calculation in which the magnetization was constrained along the $m_{z}$ axis and thus the defect states indeed split. 
This allows us to examine the effect of the orientation of magnetization on the defect state splitting.
We also found that the $m_z$-constrained magnetic configuration is 38.9 meV higher in energy (for HSE, the value is 58.4 meV) 
than the $m_{x}-m_{y}$-relaxed magnetic ground state.  
This finding suggests that the W\sub{S}-WS\sub{2} is a magnetically anisotropic material and that the easy axis lies on the $m_{x}-m_{y}$ plane. 

In Figs. \ref{fig:band} and \ref{fig:el} we show the BS and the band energies at $\Gamma$ of the $m_{z}$-constrained W\sub{S}-WS\sub{2}. 
There are six defect states for W\sub{S}-WS\sub{2} as shown in Fig. \ref{fig:el} (d). Three of these states are spin-up, and the other three are spin-down. 
For each spin species, the two degenerate states with a lower energy are composed of $d_{xy}$ and $d_{x^2-y^2}$ of the antisite W atom, and the state higher in energy 
originates from the $d_{z^2}$ orbital. It is worth mentioning that the spin-up $d_{xy}$ and $d_{x^2-y^2}$ orbitals are occupied by two unpaired electrons which 
are the source of the magnetic moment of W\sub{S}-WS\sub{2} as will be discussed in next section. 
Under SOC, the $d_{xy}$ and $d_{x^2-y^2}$ orbitals split into two bands and each of these bands is a linear combination of 
$d_{xy}$ and $d_{x^-y^2}$. For spin-up, this splitting is 296 meV, which is the highest $\Delta_{\text{SO}}$ among all the WS\sub{2} defects studied in this paper. 
For spin-down, the splitting is 87 meV. The smaller $\Delta_{\text{SO}}$ for spin-down may be related to the fact that the spin-down defect states are much higher 
in energy than the spin-up states, thus they are closer to the CBM which are the $d_{z^2}$ orbitals that do not exhibit SO splitting. 
The consequence is that the spin-down defect states are hybridized with the $d_{z^2}$ conduction bands and thus their $\Delta_{\text{SO}}$ is reduced. 
This argument is supported by the wave function analysis, which shows that both the $d_{xy}$ and $d_{x^2-y^2}$ orbitals approximately have a 
$\frac{1}{3}$ $d_{z^2}$ character. 

The $\Delta_{\text{SO}}$ from HSE+SOC are 356 meV and 62 meV for spin-up and spin-down, respectively. 
With HSE, the SO splitting of the spin-up defect states ubcreases significantly (60 meV) similar to the case of V\sub{S}-WS\sub{2}. 
However, for the spin-down defect states, with HSE the SO splitting decreases by 25 meV. 
The reason for the decreased $\Delta_{\text{SO}}$ for spin-down defect states is that HSE pushes these states further into the conduction band region, 
thereby enhancing the mixing with the $d_{z^2}$ orbitals. 
\begin{center}
	W\sub{S2}
\end{center}
W\sub{S2} is the most complicated case among the chosen defects. 
It involves ten defect states - five are spin-up and five are spin-down. As indicated in Fig. \ref{fig:el} (e), without SOC, 
the five defect states for each spin type can be categorized into three groups: two groups of doubly degenerate states which are 
lower in energy, and a single $d_{z^2}$ orbital higher in energy. 
The mixing of the conduction $d_{z^2}$ band with the 
spin-down $d_{xy}$ and $d_{x^2-y^2}$ defect bands is even worse in the case of W\sub{S2}-WS\sub{2} as the spin-down 
defect $d_{z^2}$ state is already in the conduction band region. 
The two sets of doubly degenerate states are composed of the linear combinations of the $d_{xy}$ and $d_{x^2-y^2}$ orbitals of the antisite W atom, 
and will split into four states if SOC is present. Thus, for W\sub{S2}-WS\sub{2}, there are four sets of SO splittings. 
The $\Delta_{\text{SO}}$ of each split set is 121 meV, 105 meV, 167 meV, and 138 meV, respectively, with ascending energy. 

In contrast to DFT, HSE calculation for W\sub{S2}-WS\sub{2} relaxed the magnetization onto the $m_{x}-m_{y}$ plane. 
Therefore we again constrained the magnetization along the $m_{z}$ axis. The constrained configuration is less stable than the relaxed one by 23.5 meV. 
For the magnetically constrained W\sub{S2}-WS\sub{2}, 
HSE again enchances the splittings which are not close to CBM (the first three splittings in Fig. \ref{fig:el} (e)). The increments are 46 meV, 
38 meV, 33 meV, respectively. In contrast, for the fourth splitting HSE decreases $\Delta_{\text{SO}}$ by 57 meV.  
One noteworhty feature is that the spin-up splittings are always larger than the spin-down splittings. 


\subsubsection{Magnetic moments of the W\sub{S} and W\sub{S2} antisites}
\begin{figure}
    \centering
    \begin{subfigure}{0.235\textwidth}
    \includegraphics[width=1.0\linewidth, trim =3.2cm 1cm 3.3cm 0cm, clip=true ]{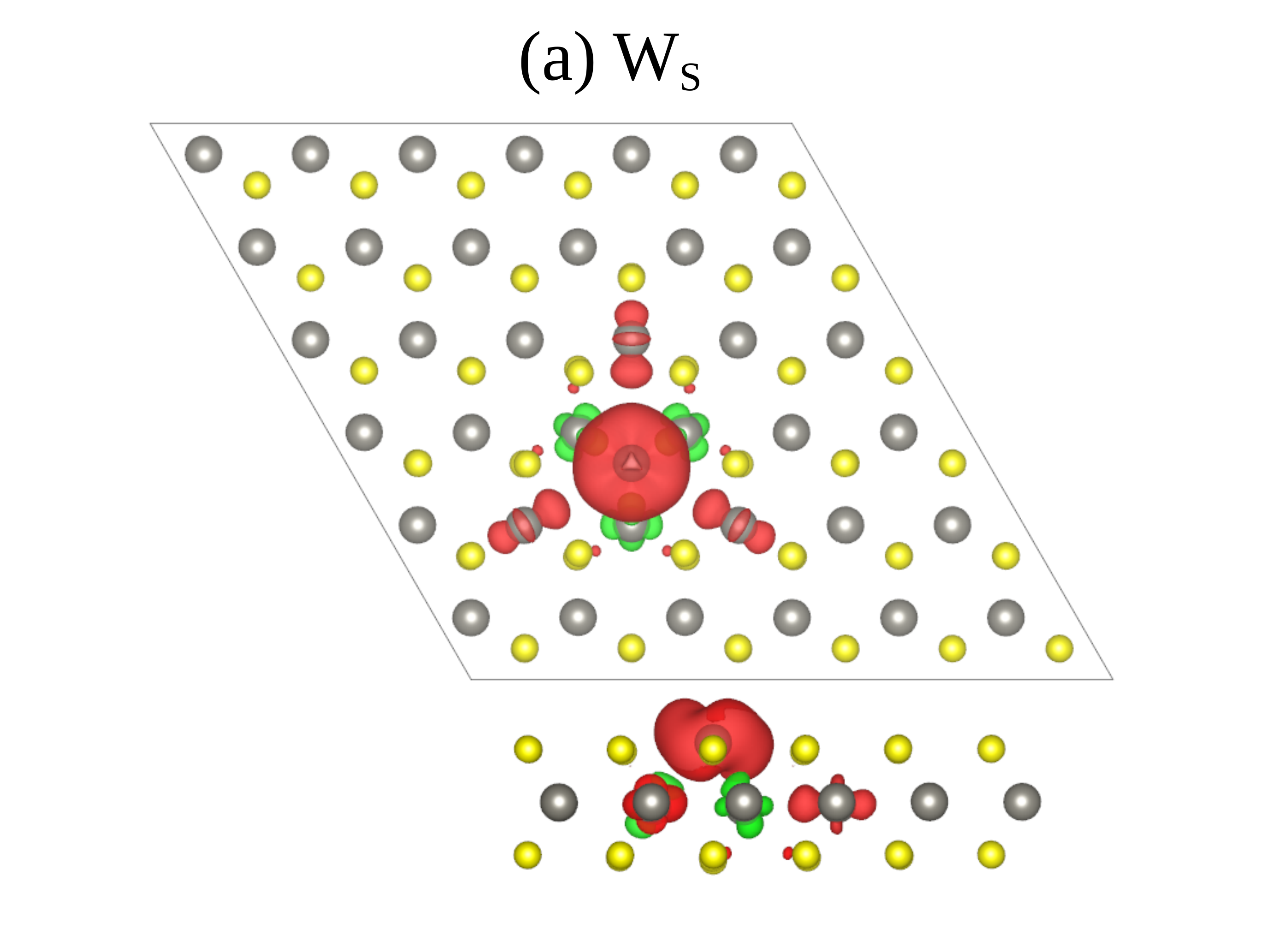}
    \end{subfigure}
    \centering
    \begin{subfigure}{0.235\textwidth}
    \includegraphics[width=1.0\linewidth, trim =3.2cm 1cm 3.3cm 0cm, clip=true ]{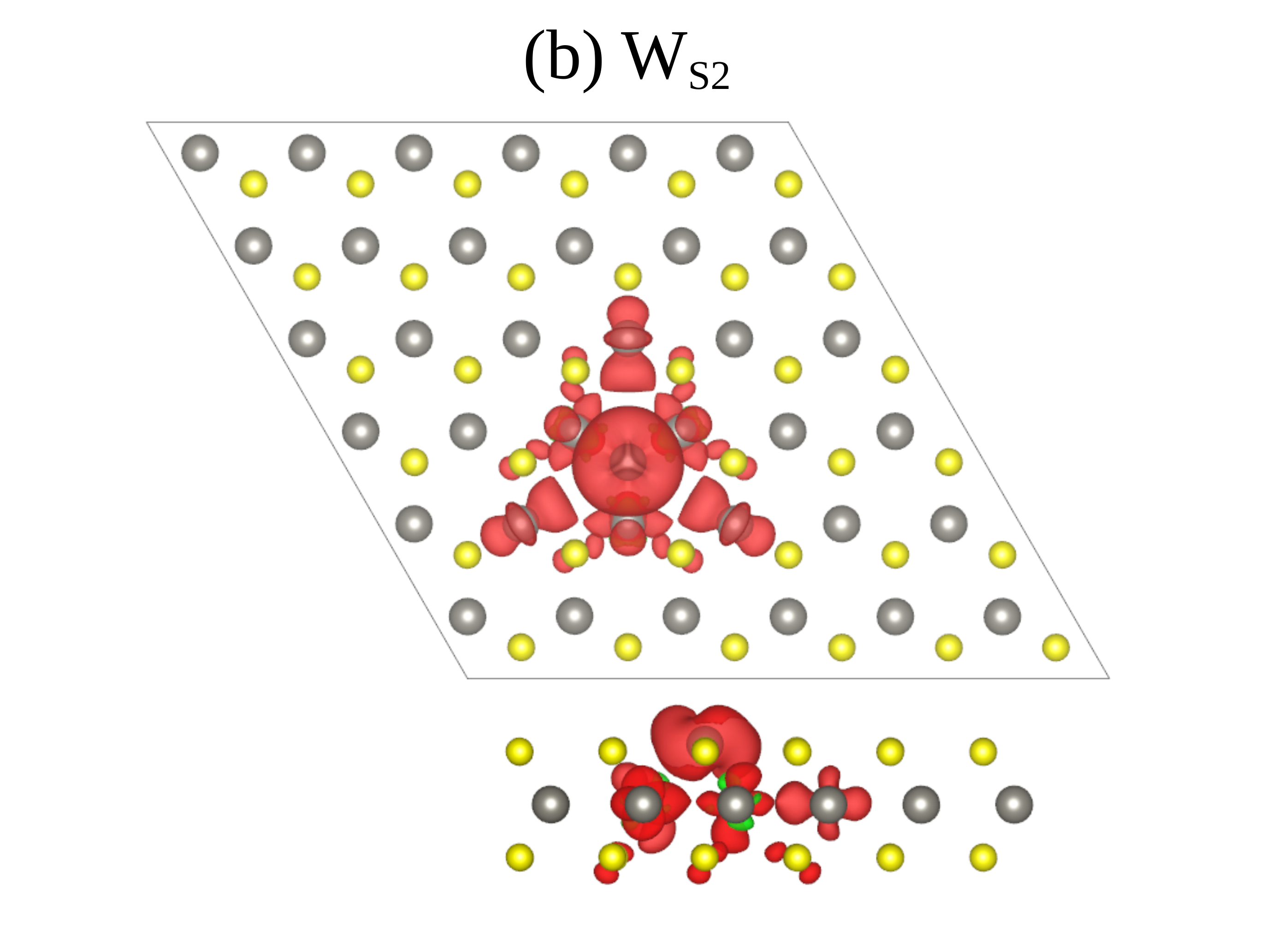}
    \end{subfigure}
    \caption{Spin density plots of (a) W\sub{S} and (b) W\sub{S2} antisites calculated by DFT.  
		The Spin-up charge density is marked in red and the spin-down density in green.
		The isosurface level is 0.002 e/\AA\sups{3}.}
    \label{fig:mag}
\end{figure}

We found that both W\sub{S} and W\sub{S2} defects possess a magnetic moment of 2 $\mu_B$. 
This is different from the result of Ref. \onlinecite{hong}, which indicated that for MoS\sub{2}, only Mo\sub{S}-MoS\sub{2} has a magnetic moment 
but not Mo\sub{S2}-MoS\sub{2}.
These magnetic moments are generated by the unpaired spin-up electrons residing on the $d_{xy}$ and $d_{x^2-y^2}$ defect states, as indicated by Fig. \ref{fig:el} 
(d) and (e). These states split under SOC. 
We defined the spin density as the difference between the spin-up charge density and 
the spin-down charge density: $\rho=\rho_{\uparrow}-\rho_{\downarrow}$ to visualize the magnetic moment distribution around the defect site. 
The resulting spin density plots are presented for both antisite defects in Fig. \ref{fig:mag}. 
At first glance, the magnetic moment seems to be fully localized on the antisite W atom, however for both W\sub{S} and W\sub{S2}, 
the $d$ orbitals of the neighboring W atoms contribute to the magnetic moment as well, and to a lesser extent also the next-nearest-neighboring (NNN) W atoms.
 are involved. For W\sub{S2}, the magnetic moment spreads to 
both the nearest-neighboring (NN) and NNN W atoms. 

We compared the ratio between the magnetic moment at the defect W atom and the total magnetic moment 
($\mu_{r}=\frac{ \mu(W_{def}) } { \mu(all) }$) to give a semi-quantitative description of the distribution of the magnetic moment. We used the VASP default 
atomic radii for W (1.455 \AA) and S (1.164 \AA) to perform the spherical integration of the spin density. 
We calculated $\mu_{r}$ using DFT (spin-polarized), DFT+SOC, and HSE+SOC methods. 
For W\sub{S}, $\mu_{r}$(DFT)= 88.4\%, $\mu_{r}$(DFT+SOC)= 88.0\%, and $\mu_{r}$(HSE+SOC)= 98\%, respectively. 
For W\sub{S2}, the corresponding values were lower at 53.1\%, 53.5\%, and 66.6\%, respectively. 
In addition, we also found that the magnetic moment distribution shown in Fig. \ref{fig:mag} has a triangular shape with a side length of around 6.4 \AA in both cases. 
Therefore these two antisite defects could also be named magnetic 'superatoms'\cite{hong}. 

Therefore one can conclude that, first, for W\sub{S} the magnetic moment is almost solely localized on the defect W atom, yet for W\sub{S2} 
the magnetic moment is centered at the defect W atom, but half of it spreads to the NN and NNN W atoms. 
Second, with the HSE hybrid functional, the magnetic moment is more localized on the dedect atom, yielding a higher $\mu_{r}$.  

\begin{figure}
	\centering
	\includegraphics[width=\linewidth, trim=2cm 0cm 4cm 3cm, clip=true]{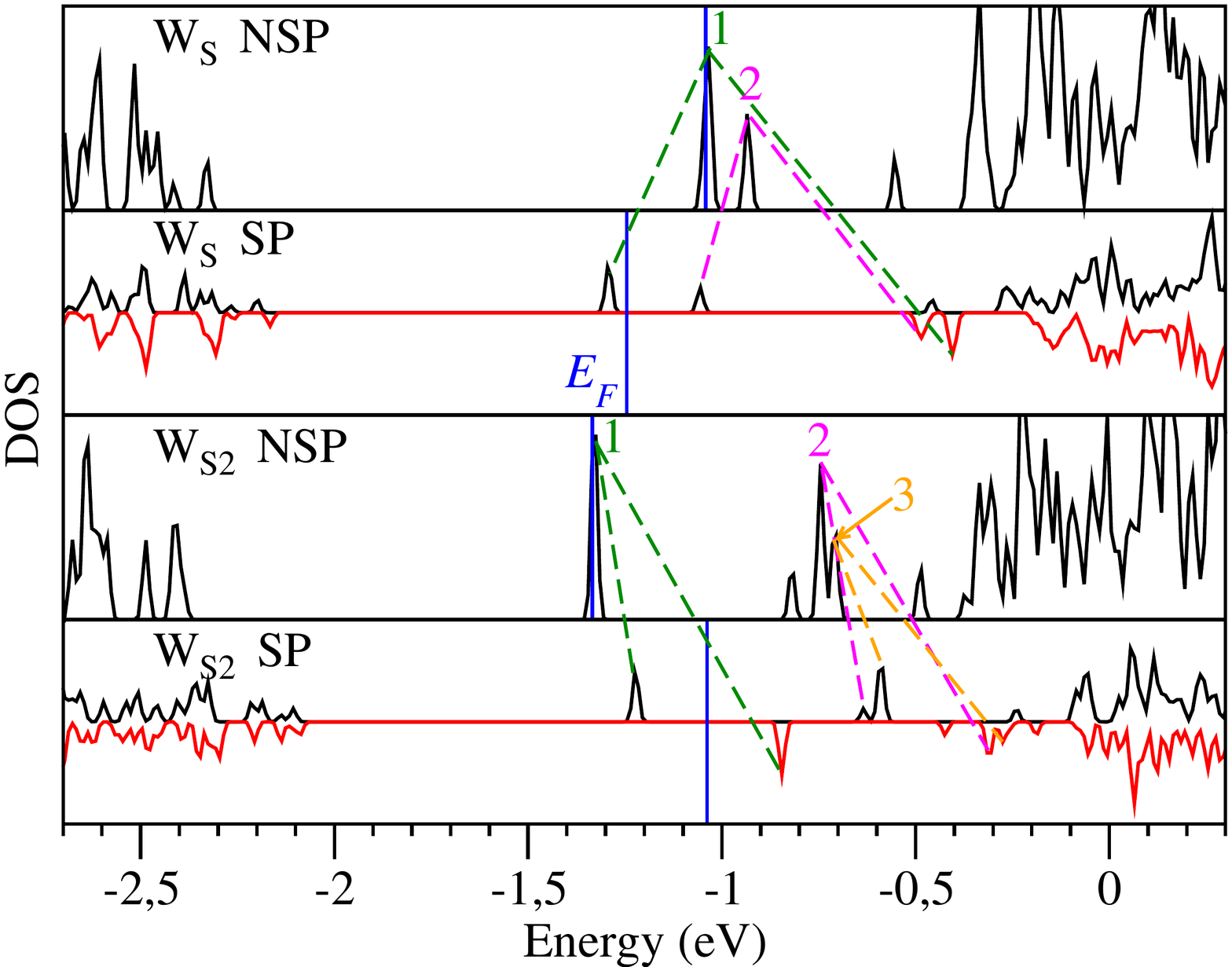}
	\caption{TDOS plots for both the non-spin-polarized (NSP) and spin-polarized (SP) W\sub{S} and W\sub{S2} antisites. The vertical blue solid lines 
			indicate the Fermi level. The colored dotted lines map the NSP $\rightarrow$ SP splitting of the defect bands}
	\label{fig:tdos}
\end{figure}

In order to trace back the origin of these magnetic moments, we compared the total energy and the density of states (DOS) 
of both the non-spin-polarized (NSP) and spin-polarized (SP) solutions of W\sub{S}-WS\sub{2} and W\sub{S2}-WS\sub{2}.
It was found that the NSP solutions are significantly higher in energy than the SP counterparts. The energy difference E(SP) - E(NSP) is 402 meV for W\sub{S}-WS\sub{2}
and 151 meV for W\sub{S2}-WS\sub{2}. Therefore both antisite configurations are indeed spin-polarized and are magnetic. 
The DOS plots of both the NSP and SP solutions for W\sub{S}-WS\sub{2} and W\sub{S2}-WS\sub{2} in Fig. \ref{fig:tdos} show clearly the magnetism. 
By combining Fig. \ref{fig:tdos}, Fig. \ref{fig:el} and the projected DOSs (PDOSs) (Fig. S2 in the SM), we performed a thorough 
eigencharacter analysis of the defect states, revealing that these states are composed of the $d$ orbitals of the antisite W atom which are numbered 
for each antisite in Fig. \ref{fig:tdos}. 
For W\sub{S}-WS\sub{2}, group 1 is composed of the $d_{xy}$ and $d_{x^2-y^2}$ orbitals and group 2 is characterized by the $d_{z^2}$ orbital. 
For W\sub{S2}-WS\sub{2} there are three groups of defect states. Group 1 and 2 are both composed of the $d_{xy}$ and $d_{x^2-y^2}$ orbitals.
However, they are now mixed with the $d_{z^2}$ orbital to different extents. Group 2 is more heavily mixed with the $d_{z^2}$ orbital than group 1. 
Group 3 is simply the $d_{z^2}$ orbital. Furthermore, for both antisite defects, only the spin-up part of peak 1 is under the Fermi level and is occupied by two 
electrons from the $d_{xy}$ and $d_{x^-y^2}$ orbitals of the antisite W atom. Therefore the magnetism and its origin is confirmed.  

\section{Conclusion}
In this study we calculated the formation energies of seven different configurations of point defects including monovacancies, interstitials and antisites. 
We found that among the point defects, V\sub{S} and S\sub{i} possess the lowest formation energies; 
$E_{f}(\text{V\sub{S}})=$1.689 eV in a W-rich chemical environment, and $E_{f}(\text{S\sub{i}})=1.211$ eV under a S-rich chemical environment.
We selected the V\sub{S}, S\sub{i}, W\sub{S} and W\sub{S2} defects to investigate the SOC band splitting of the defect states. 
We have shown that the SO splitting depends on both the orbital constitution and the orientation of magnetization of the defect states. 
The states having the $d_{xy}$ and $d_{x^2-y^2}$ character will undergo significant SO splitting when the magnetization is oriented along the $m_{z}$ magnetization axis. 
The as-generated SO splittings are 194 meV for V\sub{S}, 296 meV and 87 meV for W\sub{S}, and 121 meV, 105 meV, 171 meV, and 138 meV for W\sub{S2}. 
The hybrid functional HSE enhances the SO splitting up to 60 meV if the defect state is not close to CBM. However, it decreases the SO splitting up to 57 meV 
if the defect state is close to CBM. For S\sub{i} no SO splitting was found as the defect state is composed solely by the $d_{z^2}$ and $p_z$ orbitals. 
We also found that not only W\sub{S}, but also the W\sub{S2} defect possesses a local magnetic moment of 2 $\mu_{B}$ around the antisite W atom due to the two 
unpaired spin-up electrons occupying the $d_{xy}$ and $d_{x^2+y^2}$ defect states. The antisite W atom together with its NN and NNN W atoms thus form the so-called
superatom.  

The results presented in this Article provide new insights into the SOC behavior of the ML WS\sub{2} containing the most common point defects. 
These results are expected to be extendable to other ML MX\sub{2} systems.
In particular, the controllability of these SO split states are worth further investigation as they are highly promising in spintronics applications. 
For instance, it would be interesting to examine  whether the spins can flip when an electric field is applied. 
Also, considering the frequent occurrence of the M\sub{X2} antisites generated during the PVD synthesis of the ML MX\sub{2} membranes\cite{hong}, 
it will be interesting to increase the concentration of M\sub{X2} antisite defects and examine the interaction of the magnetic moments and their arrangement over 
space. Further development of this topic is beyond the scope of the present paper and will be addressed in future works.    
\section{Acknowledgements}
This project is financially supported by the Dutch science foundation NWO via a VIDI grant (grant number 723.012.006).  
W.F. Li acknowledges Torbj{\"{o}}rn Bj{\"{o}}rkman and Hugo Aramberri for their discussion on the SOC calculations, 
and Jyh-Pin Chou for his practical instruction on VASP settings and insight of interpreting the SOC band structures. 

\bibliography{ver5}
\end{document}